# Simulation and Visualization of Chaos in a Driven Nonlinear Pendulum – An Aid to Introducing Chaotic Systems in Physics

Louis Ehwerhemuepha[1], Godfrey E. Akpojotor[2]
[1]Computational Science, Chapman University, Orange, California, USA
[2]Theoretical and Computational Condensed Matter Physics, Physics Department, Delta State University, Abraka, Nigeria
[1]ehwer100@mail.chapman.edu: [2]akpogea@deltastate.edu.ng

**Abstract**

The presence of physical systems whose characteristics change in a seemingly erratic manner gives rise to the study of chaotic systems. The characteristics of these systems are due to their hypersensitivity to changes in initial conditions. In order to understand chaotic systems, some sort of simulation and visualization is pertinent. Consequently, in this work, we have simulated and graphically visualized chaos in a driven nonlinear pendulum as a means of introducing chaotic systems. The visualized results obtained which highlight the hypersensitivity of the pendulum to initial conditions can be used to effectively introduce the physics of chaotic system. The simulation and visualization programme is written in Python codes.

## 1. INTRODUCTION

A chaotic system may be defined as a transitive system that has a dense set of periodic points and consequently depend sensitively on initial conditions (Devaney 1992). Sensitivity dependence here means that even a microscopic change in the factors that affect the system at the onset will produce exponential changes in the overall behaviour and pattern of the system (Kellert 1993). This extreme dependence on the initial condition provides much of the unpredictability of the systems that makes it become complex. However, there is still order in this complex system because while the momentary behaviour appears random and chaotic, the general pattern of the system over time and space will display some type of order which can be modelled (Gleick, 1987).

In general, to model a system involves mapping it into an alternative system that may be simpler and can easily be manipulated and controlled yet it is described by the same mathematics (Akpojotor, 2012). Therefore to explain the above definition of a chaotic system to young undergraduates imposes on them some level of intelligence and skill in abstract mathematics. However, one can simulate and visualize physical systems in physics that can exhibit chaotic behaviour to demonstrate chaotic systems with computer programming. This is the purpose of this present study: to simulate and visualize a driven nonlinear pendulum exhibiting chaotic behaviour using a computer programme written in Python codes. The pendulum modelled in this study can be approximated to a simple pendulum when the damping and driving force quantities are removed. Thus it is the combination of the damping and driving force quantities that causes the system to transits to a chaotic system.

There are several ways or tools to simulate and visualize a physical system (Harvey et. al 2009; Few 2009). In this study, we will use a computer programme simulation to implement a model of the nonlinear pendulum in order to test it under different conditions with the objective of learning about the model's behaviour (Wolfgang. and Francisco 2007). For the visualization, a simple line plot has been employed to describe the model of the driven nonlinear pendulum used.

The mathematical model of the driven nonlinear pendulum used in this study is made of nonlinear second order differential equation which has no exact analytical solution. Consequently, the equation was solved by first breaking it into two first order differential equations and then applying difference equations. About 3700 data was obtained over a certain time evolution of the system for several different initial conditions. The result so obtained was used to introduce chaos in dynamical systems. SI units are assumed for all physical quantity.

The plan of the paper is as follows. In section 2, we will formulate the mathematical model equation for the nonlinear pendulum and then develop the Python codes for it in section 3. The results and inferences from the implementation of the programme will be discussed in section 4 and this will be followed by a summary and conclusion in section 5.

## 2. NONLINEAR DRIVEN PENDULUM

In deriving an equation of motion for the driven nonlinear pendulum, the possibility of the pendulum swinging through large angles (such that the angular displacement, theta (θ) is not necessarily equal to sin(θ)), friction as a damping force dependent on the velocity of the nonlinear pendulum and an external sinusoidal driving force was permitted. The implication is that changing the initial conditions of these parameters could alter the behaviour and pattern of the driven nonlinear pendulum.

Now putting all these parameters together, we have the equation of motion of the nonlinear pendulum (Strogatz 2004):

$$\frac{d^2\theta}{dt^2} = -\frac{g}{L}\sin\theta - q\frac{d\theta}{dt} + F_D \sin(\Omega_D t) \qquad (1)$$

where $g$ is acceleration due to gravity, $L$ is the length of string, $q\frac{d\theta}{dt}$ is the velocity dependent damping force and $F_D \sin(\Omega_D t)$ is the sinusoidal driving force. Therefore the value of $F_D$ will determine the driving force, with $F_D = 0$ being the system with no driving force.

Nonlinear equations such as Equation (1) describes many things in nature (Hoppensteadt, 2000) and are difficult, if not impossible, to solve analytically (Hirsch et al 2004). This implies that a solution does not always exist and when it does, it is not always unique.

For the pendulum considered in this study which is depicted in Fig. 1, it is easy to see that the path of the bob from A to B is an arc and therefore can be described as a circular motion. It follows then from the common textbook knowledge that the angular speed $\omega$ is given by

$$\omega = \frac{d\theta}{dt}. \qquad (2)$$

Taking Equation (2) into account in Equation (1), it becomes

$$\frac{d\omega}{dt} = -\frac{g}{L}\sin\theta - q\omega + F_D \sin(\Omega_D t). \qquad (3)$$

From Equations (2) and (3), we have the difference equation (Strogatz, 1994):

$$\omega_{i+1} = \omega_i - \frac{g}{L}\sin\theta_i \Delta t - q\omega_i \Delta t + F_D \sin(\Omega_D t)\Delta t \qquad (4)$$

with

$$\theta_{i+1} = \theta_i + \omega_{i+1}\Delta t. \qquad (5)$$

Equations (4) and (5) can then be coded and used to obtain a solution to the second order differential equation model of our system.

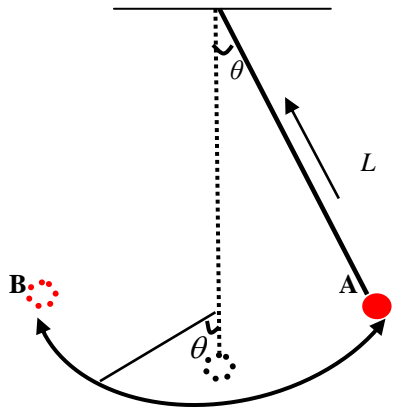

**Fig. 1** A pendulum of bob of mass m suspended from a string (of negligible mass) of length *L* which is displaced from its equilibrium position at a small angle $\theta$. Observe that the path of the bob from A to B is an arc

## 3. IMPLEMENTING THE NUMERICAL APPROACH

A modified Euler's Method for solving differential equation, Euler-Cromer Method (Strogatz 2004) was used to solve the differential equation after breaking it down to two first order systems. An excerpt of the Python codes implementing this method for our model is shown below.

```
# Implementing the Euler-Cromer Method for a set number of data points
from __future__ import division
from math import *
from scipy import *
import matplotlib.pyplot as plt

# Initial theta values
theta0 = (10*2*pi)/360
omega0 = 5*2*pi/360

# Constants
length_of_string = 9.8
gravity = 9.8
drive_frequency = 1/3
damping_force = 0.05

# Defining the driving force - controls the chaos
FD = 1.0

# Assigning the number of data points to be considered
data_points = 3700

# Preallocating space for time, theta and omega
time = zeros(data_points)
theta = zeros(data_points)
omega = zeros(data_points)

# Initializing theta and omega
theta[0] = theta0
omega[0] = omega0

# Defining time step size
dt = 0.05
```

```python
        for i in range(0, data_points-1):
            time[i+1] = time[i] + dt

            # Calculating for FD = 0, 0.1... in omegas
            omega[i+1] = omega[i] - (gravity/length_of_string)*sin(theta[i])*dt - (
                damping_force*omega[i]*dt + FD*sin(drive_frequency*time[i])*dt)

            theta[i+1] = theta[i] + omega[i+1]*dt
    plt.plot(time, theta)
    plt.ylabel("theta")
    plt.xlabel("time")
    plt.show()
```

The code excerpt contains the entire logic for implementing the numerical solution of the system in Python. It represents the case whereby the pendulum is allowed to swing unrestricted as stated above.

## 4. RESULTS AND INFERENCE

The states of the driven nonlinear pendulum system for 3700 data points is shown in Figs. 2- 6 with initial condition of $L = 9.8$, $\omega = 0$ at $t = 0$ and a damping force of 0.5. Observe that for the case of non-driving force, that is, $F_D = 0.0$, the system describes the motion of a normal damped simple pendulum depicted by the decaying sinusoidal curve as shown in Fig. 2. Here the amplitude of the system reduces due to damping with the evolution of time. In a real life scenario, this damping may be due to air drag, friction between the string and the pendulum bob and friction between the string and the anchor point.

To see the effect of how the changes in the initial conditions can affect the system, the driving force is included starting from $F_D = 0.1$. With the addition of this driving force, the system changes distinctively thereby depicting a chaotic response to these changes in a rather hypersensitive way. For the small value of $F_D = 0.1$ visualized in Fig. 3, it is observed that the earlier part of the resulting system has a chaotic damped waveform from time $t = 0 – 100$ (indicated by the red dash box) and a normal undamped sinusoidal waveform from time $t > 100$ (indicated by the green dash box). The physical explanation for the undamped region is that the driving force and the damping constant tend to cancel themselves out and this may be attributed to the driving force introducing features to cancel out the aforementioned causes of damping. While there is no report in the literature of such features yet to the best of our knowledge, the possibility to predict them with the driving force and see how they can reduce the causes of damping here is a boost to the versatility of simulation and visualization.

Now we continue to increase the driving step by *0.1* and observed the same pattern up to $F_D = 0.8$. The only changes in these new chaotic states are that the driving force seems to enhance the region of the undamped sinusoidal waveform (see Fig. 4 for the $F_D = 0.8$): while it is from $t > 100$ for $F_D = 0.1$, it is from $t > 54$ for $F_D = 0.8$. This is expected: for as explained above, the driving force seems to contribute features that reduce the effect of the causes of the damping.

It is pertinent to point out that as we increase the driving force, the undamped sinusoidal waveform becomes gradually chaotic eventually becoming abnormal at $F_D = 0.8$. Thus a further increment of *0.1*, that is, $F_D = 0.9$, produces an astonishing completely chaotic waveform as shown in Fig. 5. Observe the chaotic waveform lies above $\theta = 0$ from time $t = 0 – 65$ and then become entirely negative from time $t > 65$. Interestingly, a slight increase of the driving force by *0.1* so that $F_D = 1.0$ makes the system more chaotic as the emerging waveform which lies below $\theta = 0$ from time $t = 0 – 20$ became entirely positive for all time $t > 20$ (see Fig. 6 for the $F_D = 1.0$). This drastic change of the waveforms by slight change of the driving force by a value as small as *0.1* is analogous to the so called butterfly effect which is used to illustrate the hypersensitive nature of chaotic systems (Hilborn, 2004).

An interesting observation is that irrespective of the seemly chaotic nature of even Figs 5 – 6, the pattern of the graph for each of them is uniquely ordered. This is usually termed as orderliness in an otherwise disorderliness: as stated earlier, while momentary behaviour may appear chaotic, the general pattern over time and space can be modelled, simulated and visualized to understand the complex behaviour and probably its cause. This is one feature that has made the application of chaos to real systems possible.

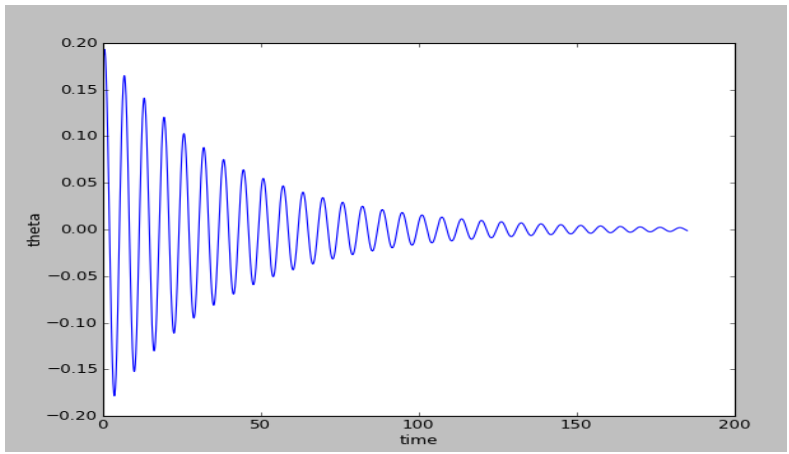

**Fig. 2** The system when the driving force is switched off, that is, $F_D = 0.0$ showing a normal damped sinusoidal waveform

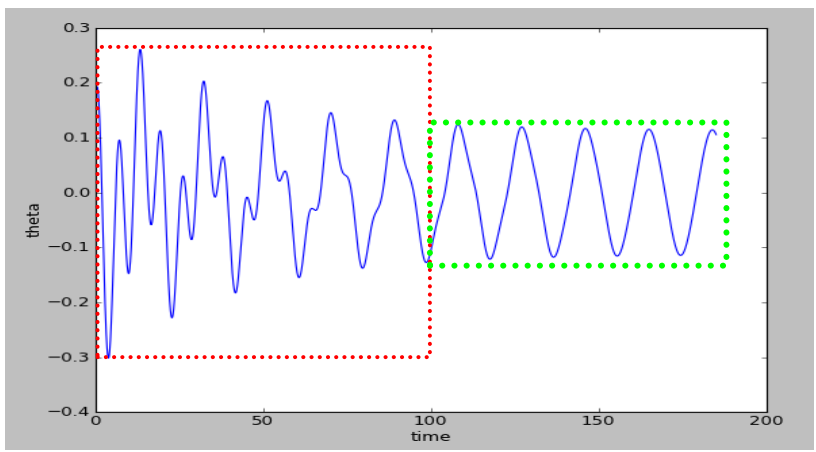

**Fig. 3** The system for varying driving force $F_D = 0.1$ showing a chaotic damped waveform from time $t = 0 – 100$, and a normal undamped sinusoidal waveform from time $t > 100$

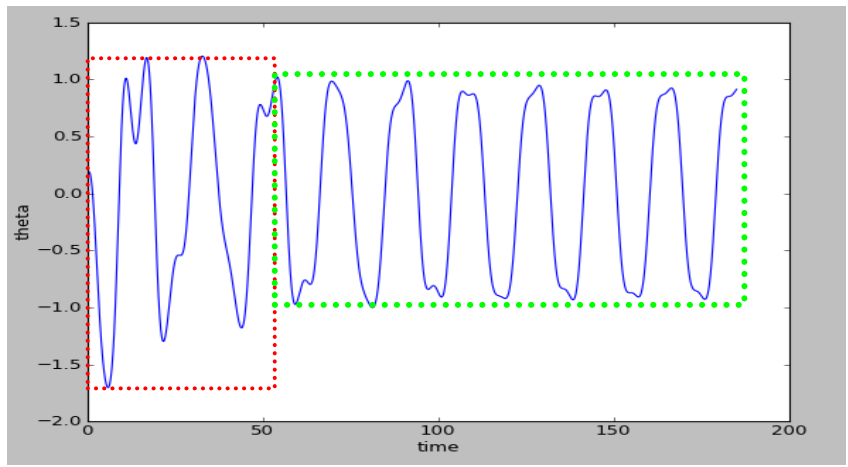

**Fig. 4** The system for varying driving force $F_D = 0.8$ showing a chaotic damped waveform from time $t = 0 –54$, and an abnormal undamped sinusoidal waveform from time $t > 54$

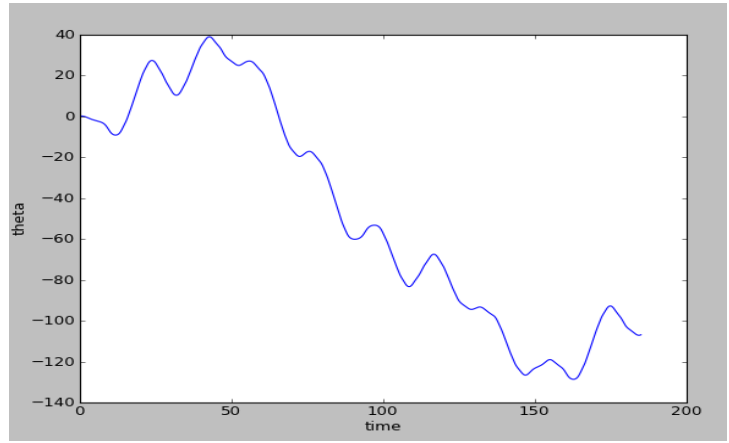

**Fig. 5** The system for varying driving force $F_D = 0.9$ showing chaotic waveform lying above $\theta = 0$ from time $t = 0 – 65$ and then became entirely negative from time $t > 65$

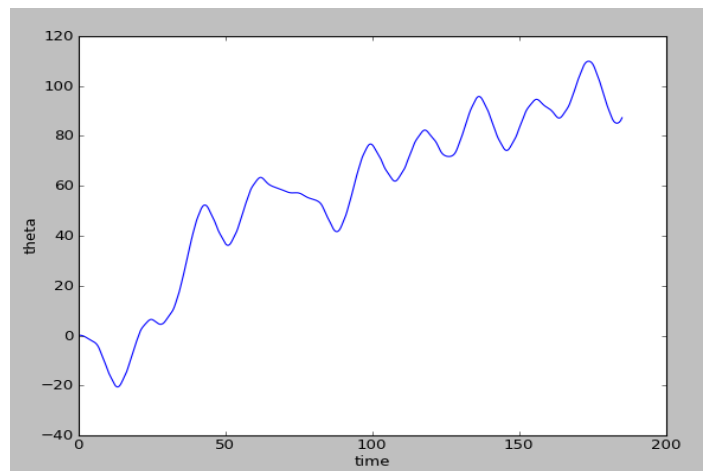

**Fig. 5** The system for varying driving force $F_D = 1.0$ showing chaotic waveform which lies below $\theta = 0$ from time $t = 0 – 20$ but became entirely positive for all time $t > 20$

## 5.    SUMMARY AND CONCLUSION

The concept of the behaviour of chaotic systems in terms of hypersensitivity to initial conditions is successfully captured by our dynamical system as it can be seen that there is a hypersensitive change in the system's behavior when the driving force is changed by just a difference of *0.1*., first from $F_D = 0$ to $F_D = 0.1$, $F_D = 0.8$ to $F_D = 0.9$ and then from $F_D = 0.9$ to $F_D = 1.0$. We have shown only these cases simply to illustrate how a change of 0.1 can affect the system and not because these were the only values of $F_D$ where chaos occurred. For as long the $F_D > 0$, the system is always chaotic, though this will also depend on the value of the damping force: systems with small damping force can be set into chaos with small driving force and vice versa. This feature is in agreement with nature as causes of damping various from one system to another. Thus the beauty of the simulation programme is that the behaviour of the system depicted by the appropriate graphical visualization can easily be seen for the various values of the driving force at a given damping force by simply varying the value of the driving force in the programme. Consequently, a wealth of information about the system that is previously unavailable can be retrieved from the visualizations developed. In general, using simulation and visualization as a teaching aid, it is easier to explain to young minds the dynamics of a chaotic systems and sustaining their interest because of the comprehensive pictorial description of the systems.

The experimental configuration of the driven nonlinear pendulum modeled in our study is difficult to set up and manage (Souza de Paula et.al., 2006). Therefore it is not straightforward to use the driven nonlinear pendulum system to demonstrate chaotic system in the laboratory. However we have demonstrated here that the application of numerical methods and visualization achieves the desired result elegantly, cost effectively and safely. The numerical solution to the differential equation used is just a series of numbers. These numbers can make no sense on their own unless visualized using an appropriate tool (such as line graphs, etc) since a sea of numbers is more of an enigma than an answer.

**ACKNOWLEDGEMENT**

We appreciate useful discussion with Myron Echenim and Famous Akpojotor. This work is partially funded by ICBR and AFAHOSITECH with grant to PACSET.